\documentclass[jap,aps,twocolumn,showpacs]{revtex4}
\usepackage{graphicx}
\begin{document}

\title{Space charge limited transport and time of flight measurements in
tetracene single crystals: a comparative study.}
\author{R.W.I. de Boer, M. Jochemsen, T.M. Klapwijk, and A.F. Morpurgo}
\affiliation{Department of Nanoscience, Faculty of Applied
Sciences, Delft University of Technology, Lorentzweg 1, 2628 CJ
Delft, the Netherlands}
\author{J. Niemax, A.K. Tripathi, and J. Pflaum}
\affiliation{3. Physikalisches Institut, University of Stuttgart,
D-70550 Stuttgart, Germany}
\date{\today}

\begin{abstract}
We report on a systematic study of electronic transport in
tetracene single crystals by means of space charge limited current
spectroscopy and time of flight measurements. Both $I$-$V$ and
time of flight measurements show that the room-temperature
effective hole-mobility reaches values close to $\mu \simeq 1$
cm$^2$/Vs and show that, within a range of temperatures, the
mobility increases with decreasing temperature. The experimental
results further allow the characterization of different aspects of
the tetracene crystals. In particular, the effects of both deep
and shallow traps are clearly visible and can be used to estimate
their densities and characteristic energies. The results presented
in this paper show that the combination of $I$-$V$ measurements
and time of flight spectroscopy is very effective in
characterizing several different aspects of electronic transport
through organic crystals.
\end{abstract}

\pacs{72.20.-i, 72.80.Le}

\maketitle

\section{Introduction}

Organic devices for electronic applications are usually based on
thin film technology \cite{Deleeuw00,Dimitrakopoulos02}. This is
particularly advantageous, as thin films of organic molecules and
polymers can be manufactured easily and cheaply in different ways.
Over the past few years, an intense research effort has resulted
in rapid improvement of the manufacturing processes, which has
allowed the commercialization of products based on organic
devices, i.e. organic electronics.

In spite of the rapid progress that has taken place on the applied
side, the rather low chemical and structural purity of the thin
films used in device fabrication has so far prevented a systematic
study of the intrinsic electronic properties of organic
semiconductors. That is because for these films, it is the defects
that determine the behavior observed experimentally
\cite{Campbell01}. As a consequence, our basic understanding of
the electronic properties of organic materials is still limited.

Improved chemical and structural purity in organic conductors can
be obtained by using single crystals of small conjugated organic
molecules. Electronic transport through single crystals of
different organic molecules has been investigated in the past by
means of time of flight (TOF) measurements \cite{Karl85,Karl99}.
It has been found that the hole mobility is approximately 1
cm$^2$/Vs at room temperature, increasing up to values in excess
of 100 cm$^2$/Vs with decreasing temperature. Since these
observations have been reported only in the highest purity
crystals, it is believed that this is the {\it intrinsic} behavior
of charge carrier mobility in organic conductors. So far however,
this intrinsic behavior has never been observed in conventional DC
transport measurements.

%%%%%%%%%%%%%%%%%%%%%% Figure growthpictures %%%%%%%%%%%%%%%%%%%%%%%%%%%%%
\begin{figure}[b]
\centering
\includegraphics[width=8.5cm]{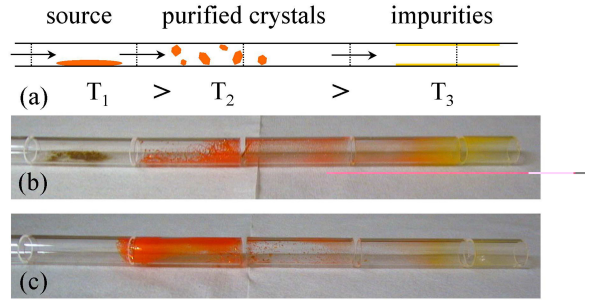}
\caption{(a) Schematic overview of tetracene crystal growth
system. Tetracene sublimes at temperature $T_1$, is transported
through the system by the carrier gas (indicated by the arrows)
and recrystallizes at temperature $T_2$. Heavy impurities (with a
vapor pressure lower that that of tetracene) remain at the
position of the source material. Light impurities (with a vapor
pressure higher than that of tetracene) condense at a lower
temperature $T_3 < T_2$, i.e. at a different position from where
the crystals grow. Therefore, the crystal growth process also
results in the purification of the material. (b) Result after
first regrowth of as-purchased tetracene. Purified tetracene
crystals are visible in the middle; the dark residue present where
the source material initially was and the light (yellow) material
visible on the right are due impurities. (c) At the end of the
second regrowth no dark residue is present at the position of the
source material, which demonstrate the purifying effect of the
growth process. \label{growthpictures}}
\end{figure}
%%%%%%%%%%%%%%%%%%%%%% Figure growthpictures %%%%%%%%%%%%%%%%%%%%%%%%%%%%%

In this paper we report an experimental study of DC transport
through tetracene single crystal and show that our results exhibit
some of the features expected for the intrinsic behavior of
organic conductors. Our investigations are based on the study of
the current-voltage ($I$-$V$) measurements in the space charge
limited current regime and on their comparison to TOF measurements
performed on identically grown crystals. As we will show, we find
overall agreement between the results obtained with the two
different methods. In particular, both $I$-$V$ and TOF
measurements show that the room-temperature effective
hole-mobility reaches values close to $\mu \simeq 1$ cm$^2$/Vs.
Both measurement techniques also show that, within a range of
temperatures, the mobility increases with decreasing temperature.
For the best samples probed by $I$-$V$ measurements this range
extends down to approximately 200 K, below which a structural
phase transition known to occur in tetracene causes a sudden drop
of the mobility. The experimental results further allow the
characterization of different aspects of the tetracene crystals.
In particular, we observe the effect of both deep and shallow
traps \endnote{We define shallow traps as traps with an energy
within a few times $k_B T$ from the edge of the valence band. Deep
traps are further separated from the valence band.}. For the
former, the measurement of the $I$-$V$ characteristics give us an
upper bound on their bulk density ($N_t^{d} < 5 \cdot 10^{13}$
cm$^{-3}$) and an estimate of their depth ($E_t^{d} \approx 700$
meV relative to the edge of the valence band). The concentration
of shallow traps is substantially larger and only a very rough
estimate can be obtained from TOF experiments.

The paper is organized as follows. We first describe the most
important aspects of the tetracene crystals growth and of the
sample preparation (section \ref{crystalgrowth}). The behavior of
the measured $I$-$V$ curves is presented in section
\ref{DCtransport}. In this section we also discuss the basic
aspects of the theoretical concepts necessary to interpret the
experimental data. Section \ref{TOF} is devoted to TOF
experiments. Finally, in section \ref{discussion} we summarize and
compare the outcome of TOF and DC transport measurements and we
present our conclusions.

\section{Crystal growth and sample preparation} \label{crystalgrowth}

Single tetracene crystals are grown by means of physical-vapor
deposition in a temperature gradient (see fig.
\ref{growthpictures}), in the presence of a stream of carrier gas,
using a set-up similar to that described in reference
\cite{Laudise98}. Both Ar and H$_2$ were used. All experiments
discussed in this paper are performed on Ar grown crystals on
which we have obtained the highest values of charge carrier
mobility. Crystal growth is performed in the dark to minimize
possible photo-activated chemical reaction of tetracene with
remnant O$_2$. These photo-induced reaction with O$_2$ are known
to occur for most polyacenes \cite{Dabestani96} and result in
chemical impurities that can act as traps for charge carriers.

%%%%%%%%%%%%%%%%%%%%%% Figure typical SCLC %%%%%%%%%%%%%%%%%%%%%%%%%%%%%
\begin{figure}[t]
\centering
\includegraphics[width=8.5cm]{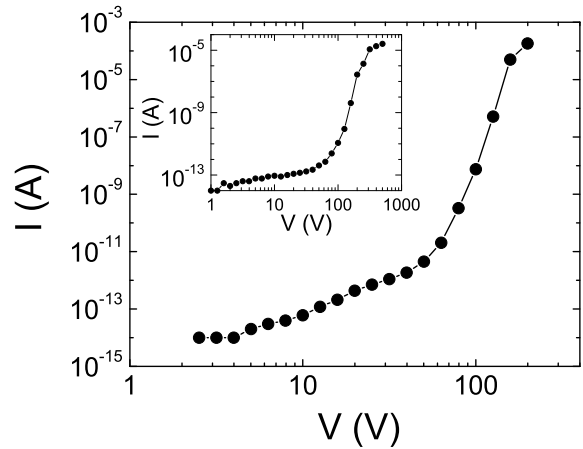}
\caption{Typical result of a DC $I$-$V$ measurement perpendicular
to the \textit{ab} plane of a tetracene single-crystal, with a
thickness $L = 30 \ \mu$m and a mobility $\mu_{\mathrm{min}}$ =
0.59 cm$^2$/Vs. The inset shows a similar measurement on a
different crystal ($L = 25 \ \mu$m, $\mu_{\mathrm{min}}$ = 0.014
cm$^2$/Vs), in which a crossing over into an approximately
quadratic dependence on voltage is visible at high voltage. In
both cases, a very steep current increase occur around of just
above $100 \ V$ that we attribute to filling of deep traps. We
observed a steep increase in current in most samples studied.
\label{typicalSCLC}}
\end{figure}
%%%%%%%%%%%%%%%%%%%%%% Figure typical SCLC %%%%%%%%%%%%%%%%%%%%%%%%%%%%%

The source material for the first crystal growth is $98 \%$ pure
tetracene purchased from Sigma-Aldrich. Crystals grown from
as-purchased tetracene are used as source material for a
subsequent growth process. For the first and second growth steps
the results of the growth processes are shown in fig.
\ref{growthpictures}. Note the large residues of impurity
molecules clearly visible after the first process (fig.
\ref{growthpictures}b) but not after the second (fig.
\ref{growthpictures}c), which directly demonstrate the usefulness
of the second regrowth process.

Tetracene crystals grown by physical-vapor deposition are large
platelets, with surfaces parallel to the {\it ab} plane. Typical
dimensions range from few square millimeters to $1 \times 1$
cm$^2$ or larger. For crystals grown by letting the growth process
proceed overnight, the typical crystal thickness ranges between
$\sim 10 \ \mu$m and $\sim 200 \ \mu$m. We have performed X-ray
structural study on a few of our thickest crystals and found a
structure consistent with known literature data
\cite{Monteath61,Holmes99}.

Multiply regrown crystals are inspected under an optical
microscope with polarized light. As it is typical for many organic
conjugated molecules, tetracene crystals are birefringent. This
allows us to select single crystals to be used for transport
experiments by choosing those samples that become uniformly dark
when changing the orientation of cross-polarizers
\cite{Vrijmoeth98}.

All $I$-$V$ and TOF measurements discussed in this paper have been
performed in the direction perpendicular to the crystal {\it ab}
plane, using electrical contacts on the two opposite faces of
crystals. Electrodes for $I$-$V$ measurements are fabricated by
connecting gold wires to the crystal surface using a
two-component, solvent-free silver epoxy, so that the epoxy is in
direct contact with the crystal. The contact area is measured
under the microscope and it is typically of the order of 0.1
mm$^2$. We use silver epoxy CW2400 (Circuitworks), which hardens
at room temperature in a few hours. This contact fabrication
method is very quick and it was chosen because it allows the
investigation of many samples in relatively short time. Other
types of contacts were tested as well, i.e. metal evaporated
contacts and colloidal graphite paint contacts, but they did not
result in any improvement of the observed electrical properties.
For TOF measurements we used silver electrodes prepared by thermal
evaporation through a shadow mask.

\section{DC transport through tetracene single crystals} \label{DCtransport}

In this section we discuss the results obtained by studying the
$I$-$V$ characteristics of approximately 100 tetracene single
crystals. We have found that the measured $I$-$V$ curves exhibit
large sample-to-sample deviations so that particular care has to
be taken in the interpretation of the experimental data. For this
reason, we first discuss how the charge carrier mobility can be
estimated using concepts of space charge limited current theory of
general validity, i.e. not sensitive to the detailed behavior of
our samples. After having presented the experimental results in
terms of the concept previously introduced, we discuss the role of
deep traps present in the bulk of the crystals and at the
metal/organic contact interface. We argue that the latter provide
the most likely explanation of the large sample to sample
variation observed in the measured $I$-$V$ characteristics.

\subsection{Estimate of the carrier mobility}
Since the band-gap of tetracene is approximately $E_g \simeq 3$ eV
\cite{Grosso}, high-purity tetracene crystals containing a
negligible amount of dopants essentially behave as insulators. It
is still possible to pass a current through tetracene crystals by
applying a sufficiently large voltage, which acts both to transfer
charge from the electrodes into the crystal and to accelerate that
charge. When the charge injected from the contacts is larger than
the charge present in the material in equilibrium the $I$-$V$
characteristics become non-linear and transport is said to occur
in the space charge limited regime \cite{Lampert70}.

For materials in which current is carried by only one carrier type
(holes in our tetracene crystals; see section \ref{TOFmeas}),
there exists an upper limit to the current that can be carried in
the space charge limited regime. This is due to electrostatics
that, at any given voltage $V$, fixes the maximum amount of charge
that can be injected into the material. For the geometry used in
our experiments, the resulting maximum current that can flow in
the space charge limited regime in the presence of an applied
voltage $V$ is \cite{Lampert70}:
\begin{equation} \label{SCLC}
I = A \frac{9 \epsilon \epsilon_0 \mu V^2}{8 L^3},
\end{equation}
with $A$ and $L$ respectively electrode area and separation and
$\epsilon$ relative dielectric constant of the material ($\epsilon
\simeq 3$ for tetracene). This upper limit is {\it intrinsic} and
it is not sensitive to any of the specific sample details that
determine the shape of the $I$-$V$ curves. For any applied
voltage, contact effects, defects or traps can only reduce the
current below the value given by equation \ref{SCLC}.

We use these considerations to obtain experimentally a lower limit
$\mu_{\mathrm{min}}$ for the mobility of charge carrier, by
measuring the current $I$ induced by a voltage $V$ and by
"inverting" equation \ref{SCLC}. We obtain:
\begin{equation} \label{SCLCinv}
\mu_{\mathrm{min}} = \frac{8 I L^3} {9 A \epsilon \epsilon_0 V^2}
\end{equation}
If $\mu_{\mathrm{min}}$ is very low - e.g., at room temperature,
much lower than the intrinsic value 1 cm$^2$/Vs typical of organic
semiconductors - this approach does not provide any useful
information. However, if the value of $\mu_{\mathrm{min}}$ is
close to 1 cm$^2$/Vs, this analysis indicates that the quality of
the crystal is high (since the intrinsic mobility $\mu >
\mu_{\mathrm{min}}$).

\subsection{Measurements of $I$-$V$ characteristics}
All the measurements of $I$-$V$ characteristics of tetracene
crystals discussed in this paper have been performed in vacuum ($P
< 10^{-5}$ mbar) and in the dark, in a two-terminal configuration.
We have used a Keithley 237 source-measure unit that permits to
apply up to 1100 V across our samples and to measure currents as
small as 10 fA. Measurements at lower temperature were performed
in the vacuum chamber of a flow cryostat. Approximately 100
samples have been measured at room temperature. Temperature
dependent measurements have been performed on most samples in
which a high value for $\mu_{\mathrm{min}}$ (0.1 cm$^2$/Vs or
better) has been found and on few of the others.

%%%%%%%%%%%%%%%%%%%%%% Figure SCLCmobilitiesoverview %%%%%%%%%%%%%%%%%%%%%%%%%%%%%
\begin{figure}[b]
\centering
\includegraphics[width=8.5cm]{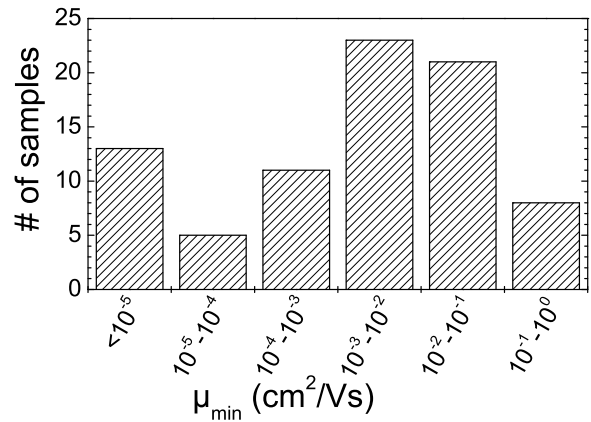}
\caption{Histogram of values for $\mu_{\mathrm{min}}$ calculated
from DC $I$-$V$ measurements performed on approximately 100
tetracene single-crystals. \label{SCLCmobilitiesoverview}}
\end{figure}
%%%%%%%%%%%%%%%%%%%%%% Figure SCLCmobilitiesoverview %%%%%%%%%%%%%%%%%%%%%%%%%%%%%

The precise shape of the $I$-$V$ characteristics measured on
different samples exhibit large deviations, whose possible origin
is discussed in the next section. Here, we focus on some important
common features observed in many of the samples investigated
($\simeq \ 100$). In particular, we often observe that in the
lower voltage range ($V < 10-100$ V, depending on the sample) the
current increases with voltage is approximately quadratic
\endnote{In samples with low $\mu_{\mathrm{min}}$ this regime may
not be visible, as the current is below the sensitivity of the
measuring apparatus.}. At higher voltage, the current increases by
many decades (typically six to eight, depending on the sample) for
a one-decade increase in voltage (fig. \ref{typicalSCLC}). In most
cases samples fail as the voltage is increased in this part of the
$I$-$V$ curve, either because too much power is dissipated through
the crystals (samples with high $\mu_{\mathrm{min}}$) or because
the electrical contacts detach from the crystal (samples with low
$\mu_{\mathrm{min}}$). In a few cases, however, we have observed
that the rapid current increase terminates by crossing over into
an approximately quadratic dependence on voltage (fig.
\ref{typicalSCLC}, inset).

%%%%%%%%%%%%%%%%%%%%%% Figure SCLCmobvstemp %%%%%%%%%%%%%%%%%%%%%%%%%%%%%
\begin{figure}[t]
\centering
\includegraphics[width=8.5cm]{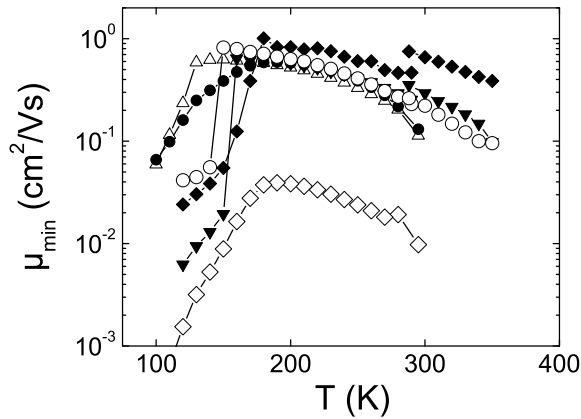}
\caption{Temperature dependence of the lower limit to the
mobility, $\mu_{\mathrm{min}}$, measured for several tetracene
single-crystals. Note the abrupt drop in mobility occurring at
different temperatures below $\simeq 180$ K, originating from a
known structural phase transition \cite{Sondermann85}.
\label{SCLCmobvstemp}}
\end{figure}
%%%%%%%%%%%%%%%%%%%%%% Figure SCLCmobvstemp %%%%%%%%%%%%%%%%%%%%%%%%%%%%%

The value of the current measured at the maximum applied voltage
is used to calculate $\mu_{\mathrm{min}}$ from equation
\ref{SCLCinv}. The calculation requires the knowledge of the
crystal thickness, which, in this geometry, corresponds to the
electrode separation $L$. The measurement of $L$ is done by
inspecting the crystals under an optical microscope. The
uncertainty of $L$ (typically $10$ to $20 \%$) is rather large and
due to the difficulty of the measurement and to the opposite
crystals surfaces not being parallel to each other. In calculating
$\mu_{\mathrm{min}}$ we have used lower estimates of $L$, in order
to be sure not to over-estimate the crystal mobility.

The histogram shown in fig. \ref{SCLCmobilitiesoverview} provides
an overview of the values of $\mu_{\mathrm{min}}$ obtained from
all measured samples. The spread in the calculated values of
$\mu_{\mathrm{min}}$ is large, as a consequence of the large
deviations observed in the measured $I$-$V$ characteristics. In
what follows we concentrate on those samples for which the lower
limit to the mobility is $\mu_{\mathrm{min}} > 0.1$ cm$^2$/Vs
\endnote{$I$-$V$ measurements performed on crystals that were
regrown three or more times do not show a statistically
significant difference from crystals that were regrown twice.}.

For most high-quality samples that did not fail during the
room-temperature measurements we have performed $I$-$V$
measurements at different temperatures. We reproducibly find that
$\mu_{\mathrm{min}}$ increases upon lowering temperature in all
the samples in which $\mu_{\mathrm{min}} > 0.1$ cm$^2$/Vs at room
temperature (fig. \ref{SCLCmobvstemp}). The same behavior has also
been observed in a couple of samples in which $\mu_{\mathrm{min}}
\simeq  0.01$ cm$^2$/Vs, although normally samples for which
$\mu_{\mathrm{min}} < 0.1$ cm$^2$/Vs at room temperature exhibit a
decrease in current as the temperature is lowered. We conclude
that samples in which the room-temperature mobility is
sufficiently close to the value of 1 cm$^2$/Vs exhibit the
behavior expected for high-quality organic semiconductors, i.e.
{\it an increasing mobility with lowering temperature}. This is
the first time that this behavior is reported in simple
two-terminal DC $I$-$V$ characteristics.

In all the high-mobility samples measured we observe that below $T
\simeq 180 \ K$, $\mu_{\mathrm{min}}$ starts to decrease when the
temperature is decrease further. In most samples, the change in
the temperature dependence of $\mu_{\mathrm{min}}$ is very sharp
(fig. \ref{SCLCmobvstemp}). This suggests that the origin of this
change is a structural phase transition, which is known to occur
in tetracene in this temperature range.

Past studies of this transition \cite{Sondermann85} have shown
that the precise transition temperature depends on details, such
as the stress induced by the adhesion between a crystal and the
substrate on which the crystal is mounted \endnote{The presence of
stress is also responsible for entire crystals breaking into small
pieces upon cooling to low temperature. The temperature at which
crystals beak can be very different for different crystals and it
seems to depend on the cooling speed.}. These studies have also
shown that the phase transition does not occur uniformly, with two
different crystalline structures coexisting in different parts of
a same crystal in a large interval of temperatures below the
transition. The coexistence of different crystal phases is
detrimental for transport, since it introduces grain boundary
junctions and regions with different bandwidth that can trap large
amounts of charge carriers. This explains the observed temperature
dependence of the mobility. The observation of the effect of this
structural phase transition on the transport properties of
tetracene provides one additional indication that the crystal
quality is high and that we are probing intrinsic effects in the
material.

\subsection{Deep traps in the bulk and at the contacts} \label{traps}
The interpretation of experimental data in terms of
$\mu_{\mathrm{min}}$ is of general validity and it does not
require any assumption regarding the sample characteristics.
Additional information can be extracted from the measurements if
one considers the behavior of the measured $I$-$V$ curves in more
detail. Here we consider the effect of deep traps present in the
bulk of the tetracene crystals and at their surface.

In general, deep traps suppress current flow by localizing charge
carriers. In the space charge limited transport regime, it can be
easily shown \cite{Lampert70} that when the applied voltage $V$ is
approximately equal to
\begin{equation} \label{Vtfl}
V_{\mathrm{TFL}} = \frac{ N_t^{d} e L^2}{ \epsilon \epsilon_0}
\end{equation}
the charge injected by the contacts is sufficient to fill all the
traps and transport occurs in the so-called trap-filled limit. At
this point (i.e., with increasing $V$ from below to above
$V_{\mathrm{TFL}}$) the measured current exhibits a large sudden
increase given by:
\begin{equation} \label{trapfillingjump}
\frac{N_v}{N_t^d} \exp{ \left ( \frac{E_t^{d}}{k_B T} \right )}
\end{equation}
In this expression, $N_v$ is the number of states in the valence
band which we take to be of the order of one state per molecule
\cite{Lampert70}.

Essentially all samples exhibit a large, steep increase in current
around a given (sample dependent) voltage (fig.
\ref{typicalSCLC}), which we interpret as due to the transition to
the trap filled limit. Using equation \ref{Vtfl} we obtain
$N_t^{d} \simeq 5 \cdot 10^{13}$ cm$^{-3}$. We find that different
samples all give comparable values. Introducing this value for
$N_t^d$ in equation \ref{trapfillingjump} we then find, taking the
magnitude of current increase measured on samples with the highest
value of $\mu_{\mathrm{min}}$, $E_t^{d} \approx 700$ meV.

The estimate of $N_t^d$ is based on the assumption, not usually
emphasized in literature, that the deep traps are uniformly
distributed throughout the entire crystal bulk. In actual samples,
due to the contact preparation process, it is likely that more
traps are present at the crystal surface under the electrodes. A
small amount of traps located close to the surface can have a
large effect in suppressing current flow. This is because charges
trapped at the surface can substantially affect the electrostatic
profile in the bulk of the crystal, which determines the current
flow in the space charge limited current regime.

To make this point more explicit, consider a $20 \ \mu$m thick
crystal in which no traps are present apart from in the first
monolayer of molecules close to the surface. Suppose that, in this
monolayer, one deep trap per every 1000 molecules is present. This
will result in a very large current suppression, as populating
these traps results in an electric field through the crystal which
corresponds, in the case considered, to approximately 1000 Volts
applied between the electrodes. For an $I$-$V$ measurement this
would imply that, as the voltage across the electrodes is
increased, the surface traps are initially filled and no current
flows until more than 1000 Volts are applied. For comparison, for
a $20 \ \mu$m thick crystal with no surface traps and a bulk
density of traps of $5 \cdot 10^{13}$ cm$^{-3}$ transport already
occurs in the trap free limit when 200 V are applied across it.
For this reason, our estimate of $N_t^d$ is a higher limit to the
bulk density of deep traps.

The strong sensitivity of the $I$-$V$ curves to traps located at
the metal/organic interface makes these traps a logical
explanation for the large sample to sample variation observed in
the experiments. Evidence for the relevance of contact effects is
provided by the rather good sample-to-sample reproducibility
observed in TOF measurements (see section \ref{TOFmeas}) as
compared to DC transport measurements.

\section{TOF experiments on tetracene single crystals} \label{TOF}

\subsection{Technical aspects of TOF experiments} \label{TOFsetup}
The time of flight (TOF) spectroscopy is based on two fundamental
steps. First, by photo-excitation electron hole pairs are
generated near the crystal surface. Second, in the applied
electric field the charge carriers move to the electrodes and the
corresponding displacement current is measured \cite{Grosso}.
Therefore, studies on the transport behavior are free of contact
effects and the technique is selective on the respective type of
charge carrier by choice of the polarity of the external voltage.

For the TOF measurements the tetracene crystals are covered on
both sides by a thin layer of Ag thermally evaporated and,
afterwards, mounted on a Cu support acting as back-electrode as
well as thermal contact for the temperature dependent studies. The
$15-20$ nm thick Ag contacts showing almost bulk conductivity but
are still sufficiently transparent for photo-induced generation of
charge carriers at the front electrode. As light source for the
charge carrier generation, a nitrogen laser in single shot mode
($\lambda$ = 337 nm, pulse width 0.76 ns) is used. The suitability
of the used wavelength for photo-excitation has been proven by
UV-VIS absorption measurements on the crystals. In addition, from
the TOF pulse shape and the absorption spectra we can conclude
that charge carrier generation takes place in the topmost fraction
of the crystals mainly, i.e. the depth of charge carrier
generation (several $\mu$m) can be neglected with respect to the
thickness of the crystals ($\sim 100 \ \mu$m).

%%%%%%%%%%%%%%%%%%%%%% Figure Stuttgart1 %%%%%%%%%%%%%%%%%%%%%%%%%%%%%
\begin{figure}[b]
\centering
\includegraphics[width=8.5cm]{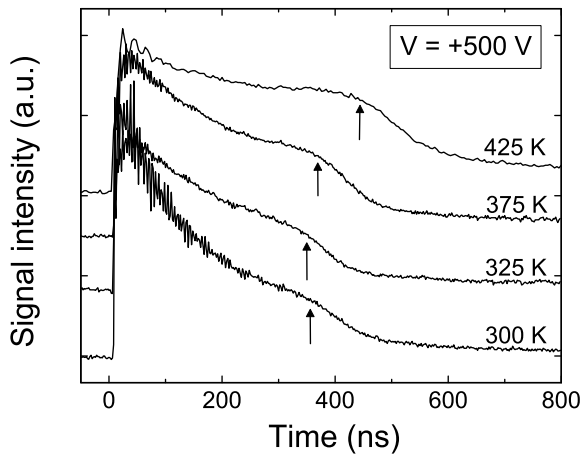}
\caption{Hole TOF pulses measured at different temperatures in the
range from room temperature and 150 $^{\circ}$C. The applied
voltage is +500 V. The arrows point to the transit times.
\label{Stuttgart1}}
\end{figure}
%%%%%%%%%%%%%%%%%%%%%% Figure Stuttgart1 %%%%%%%%%%%%%%%%%%%%%%%%%%%%%

The displacement current is measured as a voltage drop across a
resistor connected in parallel to the crystal. The resistor is
chosen in such a way that the time constant of the RC-part is much
smaller than the transit time $\tau$. TOF studies in the range
from room temperature up to 450 K were carried out in a heating
stage at ambient pressure, the upper temperature limit caused by
sublimation of tetracene at around 450 K \cite{Cox70}.

\subsection{Measurement of TOF transients} \label{TOFmeas}
Temperature dependent TOF measurements have been performed on
three different single crystals giving essentially identical
results. The room-temperature mobility values obtained from the
different crystals were very close to each other (ranging from
$0.5$ cm$^2$/Vs to $0.8$ cm$^2$/Vs) and their temperature
dependence was qualitatively identical. This is in net contrast
with the large sample-to-sample variations observed in DC
transport measurements. From the outcome of the TOF experiments we
conclude that the quality of the grown crystals is rather
reproducible and that the sample to sample variations observed in
the DC $I$-$V$ originates from irreproducibility in the contact
preparation
\endnote{Additional evidence for contact effects is given by the
asymmetry in the $I$-$V$ curves and polarity dependent hysteresis
often observed in samples with low $\mu_{\mathrm{min}}$.}. This is
consistent with the fact that TOF experiments are not very
sensitive to the contact quality whereas DC $I$-$V$ measurements
are.

Representative TOF pulses for positive charge carriers measured at
various temperatures in one of the three tetracene single crystals
are shown in fig. \ref{Stuttgart1}, from which the transit time
$\tau$ can be easily extracted. In contrast to these well-defined
TOF pulses measured for hole transport, only dispersive transport
is observed for electrons throughout the measured temperature
range. This indicates strong trapping for electrons, which is why
in section \ref{DCtransport} we have used single carrier space
charge limited current theory to interpret the behavior of the
measured $I$-$V$ curves.

\begin{figure}[thb]
\centering
\includegraphics[width=8.5cm]{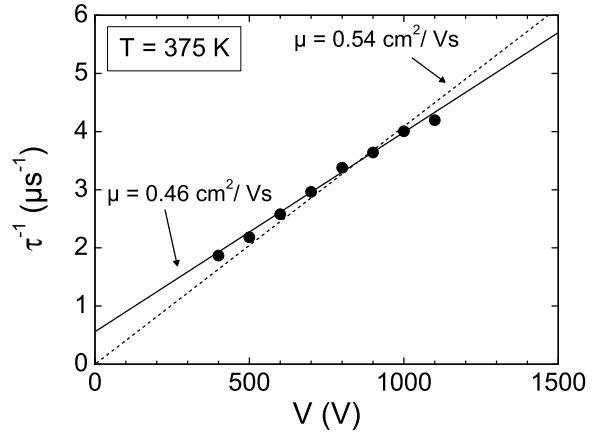}
\caption{Transit time of hole TOF pulses versus applied voltage
estimated at 375 K showing a linear (Ohmic) dependence. The
continuous and dotted lines represents, respectively, the best
linear fit and the best linear fit passing through the origin. The
corresponding difference in the $\mu$ values gives a measure of
the uncertainty in the extracted mobility. \label{Stuttgart2}}
\end{figure}

Assuming a constant electric field \textit{E} across the crystal,
the mobility of the holes is related to the transit time and to
the crystal thickness $L$ by:
\begin{equation}
\mu = \frac{L}{E \tau}
 \label{TOFmobility}
\end{equation}
For the values of electric field used in our studies $\tau^{-1}$
depends linearly on $E$ (see fig. \ref{Stuttgart2}) so that $\mu$
does not depend on electric field. All the mobility values
discussed in this paper have been estimated from this ohmic regime
only.

%%%%%%%%%%%%%%%%%%%%%% Figure Stuttgart2 %%%%%%%%%%%%%%%%%%%%%%%%%%%%%
%%%%%%%%%%%%%%%%%%%%%% Figure Stuttgart2 %%%%%%%%%%%%%%%%%%%%%%%%%%%%%

\begin{figure}[t]
\centering
\includegraphics[width=8.5cm]{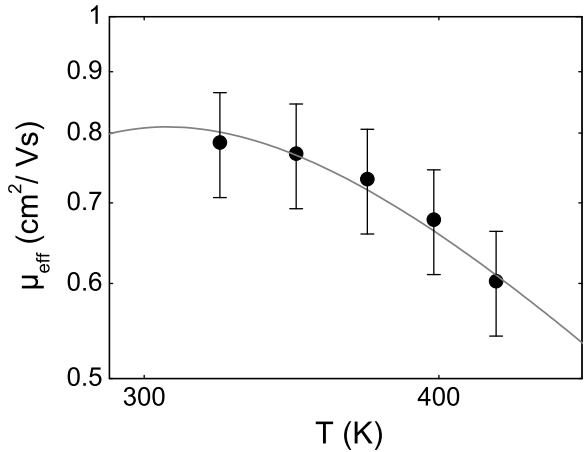}
\caption{Temperature dependence of the hole mobility (full
circles). The fit (straight line) is described by equation
\ref{mueff} using an exponent $n = 2$, a mobility of 1.4
cm$^2$/Vs, a trap energy of 0.13 eV and a density ratio between
shallow traps and tetracene molecules of $5 \cdot 10^{-3}$.
\label{Stuttgart3}}
\end{figure}

In the absence of traps equation \ref{TOFmobility} represents the
intrinsic mobility $\mu$ of the organic material, which typically
varies as an inverse power of temperature, i.e. $\mu \propto
T^{-n}$ ($n \simeq \ 2-3$ depending on the specific organic
molecules \cite{Grosso}). If shallow traps \endnote{The impact of
a low density of deep traps ($E_t^d \ \gg \ k_BT$) is of minor
importance for TOF experiments, contrary to the case of $I$-$V$
measurements, since for these traps the relaxation time is much
longer than the transit time and hole captured by deep traps do
not give any signal.} are present, however, the measured mobility
is just an "effective" mobility $\mu_{\mathrm{eff}}$ related to
the $\mu$ by \cite{Grosso}:
\begin{equation}
\mu_{\mathrm{eff}}(T) = \frac{\mu(T)}{1 + \frac{N_t^{s}}{N_v}
\left [ \exp \left ( \frac{E_t^s}{k_B T} \right )-1 \right ]}
 \label{mueff}
\end{equation}
where $N_t^{s}/N_v$ is the density ratio between shallow traps and
organic molecules in the crystal and it is assumed that all the
shallow traps have the same energy depth $E_t^s$ relative to the
valence band. A plot of the measured $\mu_{\mathrm{eff}}$ vs. $T$
is shown in fig. \ref{Stuttgart3}. The saturation with lowering
temperature is a characteristic signature of shallow traps.

The effect of shallow traps is not only visible in the temperature
dependence of the mobility, i.e. of the transit time of TOF
pulses, but also in their shape. Specifically, fig.
\ref{Stuttgart1} shows that the signal intensity increases with
increasing temperature and that the pulse becomes more
rectangular. Both effects are due to reduced trapping at elevated
temperature, which make the sample behave more closely to the
ideal trap-free condition.

%%%%%%%%%%%%%%%%%%%%% Figure Stuttgart3 %%%%%%%%%%%%%%%%%%%%%%%%%%%%%
%%%%%%%%%%%%%%%%%%%%%% Figure Stuttgart3 %%%%%%%%%%%%%%%%%%%%%%%%%%%%%

We have attempted to use equation \ref{mueff} to fit the measured
temperature dependence of $\mu_{\mathrm{eff}}$ and to determine
the values of $N_t^s$ and $E_t^s$. However, the limited
temperature range which is experimentally accessible, the unknown
intrinsic room-temperature mobility of tetracene and of the value
$n$ determining its temperature dependence make it impossible to
determine trap density and concentration precisely. Even when we
set $\mu$(300 K) = 1 cm$^2$/Vs and $n = 2$, we find that different
combinations of $N_t^s$ and $E_t^s$ produce a satisfactory fit of
our data. From this analysis we can however roughly estimate that
$N_t^s \approx 10^{18}$ cm$^{-3}$ and $E_t^{s} \approx 100$ meV. A
more precise determination of these parameters would require to
extend the temperature dependent measurements over a wider
temperature range. In practice, however, the temperature range is
limited by sublimation of tetracene ($T > 450$ K) on the high end
and by the blurring of the TOF pulse on the low end ($T < 300$ K).

\section{Summary} \label{discussion}

This is the first time that systematic DC transport measurements
and TOF experiments have been performed on identically grown {\it
single crystals} of organic molecules and it is interesting to
find overall agreement in the comparison of the results obtained
with the two techniques. Specifically, from both measurement
techniques we conclude that the room temperature hole mobility is
close to 1 cm$^2$/Vs and that its temperature dependence is
non-monotonic \endnote{Although the general trend in the $T$
dependence of the mobility is similar in both the $I$-$V$ and TOF
experiments, the temperature at which the mobility reaches its
optimum value is different. We suggest that the cause of this
difference is that, in the case of $I$-$V$ measurements, a much
larger amount of charge is being injected, so that more (shallow)
traps are filled.}. For TOF measurements on tetracene, this
behavior had been observed previously \cite{Berrehar76}. On the
contrary, an increase in mobility with lowering temperature
observed by measuring the DC $I$-$V$ curves in the space charge
limited transport regime had not been reported previously neither
for tetracene nor, to the best of our knowledge, for any organic
un-doped semiconductor. This observation, together with the
observed effect of the structural phase transition on the hole
mobility, indicate that signatures of the intrinsic electronic
properties are visible in the $I$-$V$ measurements, which
demonstrates the high quality of the crystals.

In spite of the crystal quality, the effect of imperfections is
still clearly visible. In particular, $I$-$V$ and TOF measurements
provide complementary information about the presence of deep and
shallow traps. $I$-$V$ measurements allow us to infer an upper
limit for the bulk density of deep traps and their activation
energy (see section \ref{traps}). For shallow traps the
observation of a maximum in mobility in TOF measurements around
room temperature indicates that the typical activation energy is
$\sim 100$ meV, but a more precise value as well as a reliable
estimate of the density cannot be presently obtained. We believe
that these shallow traps originate from local deformations of the
crystals, such as those due to mechanical stress or to
electrically-inactive chemical impurities as well as from chemical
impurities interacting only weakly with the surrounding host
molecules. Within a conventional band picture it is easy to see
that these deformations would have an important effect, as they
can change locally the band gap of tetracene and form spatially
localized "pockets" of holes. This mechanism could account for a
fairly large density of shallow traps and for a trapping energy of
$\sim 100$ meV \cite{Silinsh94}, which corresponds to only a few
percent change in the 3 eV tetracene gap.

Not only the similarities but also the differences between $I$-$V$
and TOF measurements provide useful information. Particularly
noticeable is the reproducibility of TOF measurements in contrast
to the large sample to sample deviations observed in the $I$-$V$
characteristics. Three out of three crystals studied by means of
TOF gave $\mu \simeq 0.5-0.8$ cm$^2$/Vs, whereas out of
approximately 100 sample measured only 5 to 10\% gave a mobility
larger than 0.1 cm$^2$/Vs. The reproducibility obtained in TOF
measurements indicate that different tetracene crystals grown in
our set-up exhibit only minor differences in their properties and
that these differences cannot account for the large spread of
$I$-$V$ characteristics observed experimentally.

We conclude that the large sample to sample variations observed in
the measurement of $I$-$V$ characteristics mainly originate from
the quality of the electrical contacts. As mentioned above, this
is critically important for $I$-$V$ measurements, but not for TOF
measurements. Since the large values of $\mu_{\mathrm{min}}$
obtained in the best samples indicate that it is possible to
fabricate "high-quality" contacts using silver epoxy, we infer
that the effects determining the contact quality in our samples
are mainly of {\it extrinsic} nature. Our estimates suggest that
deep-traps present under the contacts at the crystal surface
(introduced during the contact fabrication) play an important
role. This is because even a very small surface density of these
traps can substantially perturb the electrostatic profile in the
crystal bulk that determines the current flow in the space charge
limited transport regime.

\section{Acknowledgments}
We gratefully acknowledge N. Karl for his invaluable help and
support. This work was financially supported by FOM and by
DFG-Schwerpunkt (project No. KA 427/8). The work of AFM is part of
the NWO Vernieuwingsimpuls 2000 program.

\end{document}